\documentclass[aps,prl,twocolumn,noshowpacs,superscriptaddress,groupedaddress]{revtex4}  % for review and submission
\usepackage{graphicx}  % needed for figures
\usepackage[FIGTOPCAP,bf]{subfigure} % needed for sub figures
\usepackage{dcolumn}   % needed for some tables
\usepackage{bm}        % for math
\usepackage{amssymb}   % for math
\usepackage{balance}  % for  \balance command ON LAST PAGE  (only there!)
\usepackage{epstopdf}
\usepackage{color}
\begin{document}
\title{Emergent oscillations assist obstacle negotiation during ant cooperative transport}

\author{A. Gelblum} \affiliation{Department of Physics of Complex Systems, Weizmann Institute of Science, Rehovot 7610001, Israel.}\author{I. Pinkoviezky}\affiliation{Department of Chemical Physics, Weizmann Institute of Science, Rehovot 7610001, Israel.} \author{E. Fonio}\affiliation{Department of Physics of Complex Systems, Weizmann Institute of Science, Rehovot 7610001, Israel.} \author{N.S. Gov}\affiliation{Department of Chemical Physics, Weizmann Institute of Science, Rehovot 7610001, Israel.} \author{O. Feinerman}\affiliation{Department of Physics of Complex Systems, Weizmann Institute of Science, Rehovot 7610001, Israel.}

\begin{abstract}
\textcolor{black}{Collective motion by animal groups is affected by internal interactions, external constraints and the influx of information. A quantitative understanding of how these different factors give rise to different modes of collective motion is, at present, lacking.} Here, we study how ants that cooperatively transport a large food item react to an obstacle blocking their path. Combining experiments with a statistical physics model of mechanically coupled active agents, we show that the constraint induces a deterministic collective oscillatory mode that facilitates obstacle circumvention. We provide direct experimental evidence, backed by theory, that this motion is an emergent group effect that does not require any behavioral changes at the individual level. We trace these relaxation oscillations to the interplay between two forces; informed ants pull the load towards the nest while uninformed ants contribute to the motion's persistence along the tangential direction. The model's predictions that oscillations appear above a critical system size, that the group can spontaneously transition into its ordered phase, and that the system can exhibit complete rotations are all verified experimentally. We expect that similar oscillatory modes emerge in collective motion scenarios where the structure of the environment imposes conflicts between individually held information and the group's tendency for cohesiveness.
\end{abstract}

\maketitle

\begin{figure}
\includegraphics{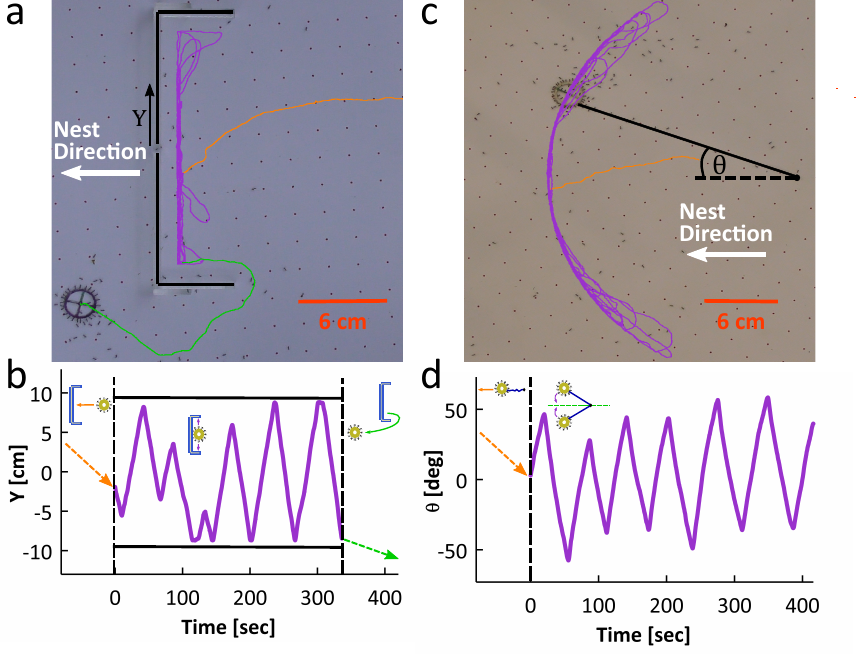}
\caption{Examples of oscillatory motion under constrained conditions - \textbf{(a,c)} The motion of cooperatively carrying ants was constrained by either a U-shaped barrier (black lines) (a) or a thin tether (black line) (c). The trajectory of the load is overlaid on both images. Different colors represent the three stages of motion: free motion before encountering constraint (orange), oscillations near block (a) or when string is fully taut (c) (purple), and motion towards nest following obstacle circumvention (green, appears only in (a)). \textbf{(b,d)} Time series of the load's motion. Black dashed lines separate the three stages of motion described above. Approaching and post-problem solution phases are represented by dashed arrows (orange and green, respectively). Oscillations are plotted in purple. In (b) ${\mathrm{Y}}$ is a coordinate defined from the center of the barrier, as portrayed in (a), and barrier edges are marked by solid black lines. The angle, $\theta$, in (d) is measured relative to the direction of the nest (c).}
\end{figure}

\fbox{\begin{minipage}{23em}

\section*{Significance}

When ants that cooperatively carry large food items encounter obstacles, they switch their collective motion from radial, nest-bound movement to nearly deterministic, tangential oscillations which facilitate obstacle circumvention. This oscillatory motion \emph{cannot} be explained by a "wisdom-of-the-crowds" model in which all ants are informed of the direction to the nest and the collective motion reflects an average effort. Rather, oscillations emerge due to two, often conflicting, forces: some ants pull the load  towards the nest while others align their pull with the momentary direction of motion. Interestingly, these two forces suffice to generate both nest-bound and oscillatory motion without any requirement that individual ants directly sense the obstacle. This is an example of emergent problem solving at the group-level scale.
\end{minipage}}

\textcolor{black}{The collective motion of animal groups is affected by several factors. First, the tendency for global alignment and cohesiveness \cite{cavagna2010scale,ariel2015locust,vicsek2012collective} which are often the result of near-neighbor interactions \cite{ballerini2008interaction,ariel2015locust,gautrais2012deciphering,herbert2011inferring,harpaz2014receptive}. Quantitative variations in these local interactions can lead to qualitatively different global modes of collective motion \cite{couzin2002collective}. Second, The motion of a group is affected by influential leaders, which bring new knowledge into the system \cite{couzin2011uninformed,stroeymeyt2011knowledgeable,brown2002social,ward2011fast,krause2000leadership,biro2006compromise,gelblum2015ant}. Last, the collective motion has to comply with environmental constraints such as boundaries or obstacles \cite{buhl2006disorder,tunstrom2013collective}. When analyzing group motion, it is the interaction between these factors that must be considered \cite{biro2006compromise,miller2013both,couzin2011uninformed}. Observations suggest that such interactions could trigger transitions between different global modes of collective motion \cite{tunstrom2013collective}. A quantitative understanding of such collective behavioral shifts is still lacking.}

\textcolor{black}{Here, we approach this question in the framework of cooperative transport \cite{czaczkes2013cooperative,mccreery2014cooperative,berman2011experimental} by \emph{Paratrechina longicornis} ants \cite{trager1984revision,czaczkes2013prey}. This behavior occurs as several ants coordinate their forces to collectively carry food items that are too big and heavy for any single ant.} %Cooperative transport by \emph{P. longicornis} occurs during the summer as the ants collectively carry large arthropods to their nest.}

\textcolor{black}{We have previously suggested that to efficiently move towards their nest, the ants balance between the well-coordinated pull of non-informed individuals and directional information brought in by informed leaders \cite{gelblum2015ant}. However, these two forces are typically aligned and this obscures their relative effects. In this work, we use an external constraint to decouple the effect of informed and non-informed individuals and expose the interplay between them. Surprisingly, we find that this manipulation transitions the group into a new collective mode of oscillatory motion. We combine experimental manipulations and theory to show that these oscillations cannot be attributed to individual ants  but are, rather, an emergent result of the internal conflicts between the different forces acting on this system.
}

\section*{Results}

\subsection*{Oscillatory motion in the vicinity of obstacles}
\begin{figure}
\includegraphics[scale=1]{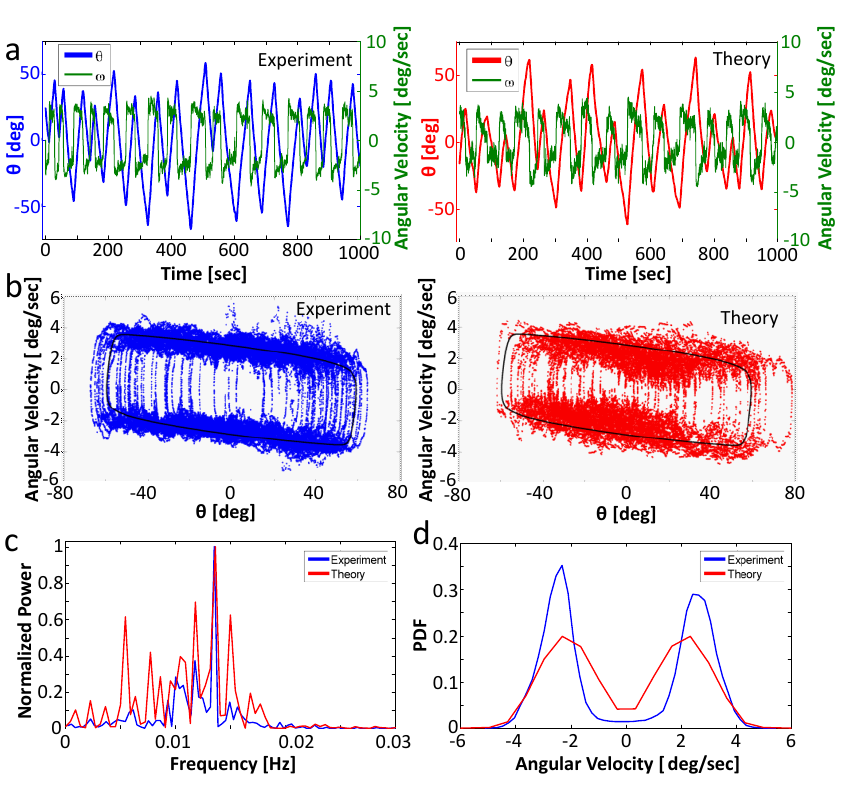}
\caption{Pendular motion characteristics in experiment (blue lines) and theory (red lines). \textbf{(a)} Time series of angle around nest direction ($\theta$) and corresponding angular velocity (green), for a 1 cm radius load tethered by an 18 cm string. \textbf{(b)} Phase space of oscillation data, such as shown in (a); black line depicts the analytic result for the limit cycle obtained from the simplified 1D model (see text below, Eqs.%\ref{soe}
1-2). \textbf{(c)} Comparison of experimental and theoretical normalized power spectral density of the angular time series presented in (a). \textbf{(d)} Experimental and theoretical distributions of angular velocity. \textcolor{black}{The experimental data presented in this panel represents $98$ minutes of oscillatory motion}.}
\end{figure}

When cooperatively carrying ants reach a barrier they commence a back and forth perpendicular motion between its edges that, eventually, leads the group around the obstacle (figure 1a,b). This global phenomenon can stem from two distinct types of individual behavior. Individual carrying ants may sense the obstruction and explicitly change their mode of motion. Alternatively, single ant behavior may remain unchanged such that the observed collective motion upon reaching an obstacle naturally emerges from the interactions between ants.

To distinguish between these two possibilities, we allowed groups of ants carry a load tethered by a thin string.  This configuration captures the situation encountered by the group when approaching a slightly raised obstacle, such as a twig, that allows a free underpass for the ants but is too low for the large load to pass (see figure SI 1) The tether configuration is useful in two ways. First, contrary to a barrier which individual ants may sense directly, the string constrains the motion in a way that is not apparent to individual ants (\textit{i.e.}, there is no simple way to distinguish string tension from forces applied by other ants). Second, while this configuration reproduces the effect of a barrier on the global motion pattern (figure 1c), the tether does not permit escape. This leads to long-lived non-linear oscillations (figure 1d and SI movie S1), which we filmed and tracked using a specialized imaging software that extracts the load center-of-mass as well as the number of carrying ants (see Section SI 1.1). This provided large data-sets that allow for accurate statistics of the collective motion (figure 2).

The angular time series of a tethered load exhibits relaxation oscillations \cite{strogatz2014nonlinear} (figure 2a-b) centered around the direction of the nest (figure 1c). We find a distinct frequency for the oscillations (figure 2c), which implies an underlying deterministic motion. Remarkably, these oscillations occur in the midst of intense stochastic activity; new ants constantly arrive from the nest, carriers attach and detach from the load, and ants adjust their carrying roles and pulling directions \cite{gelblum2015ant} (see SI Movie S1).

\subsection*{Microscopic model supports emergent oscillations}
{\color{black} To understand the origin of this oscillatory behavior we used our theoretical model for cooperative transport which is based on detailed experimental observations \cite{gelblum2015ant}. Fitting this model to experimental results holds a double advantage. First, the model was developed to describe  collective motion in the absence of obstacles; Comparing its solutions in the presence of a constraint  with the experimental measurements therefore provides a stringent test to the model's assumptions.  Second, since the model does not describe obstacles it does not allow ants to explicitly shift their behavioral program upon reaching one. Therefore, if the constrained version of the model exhibits oscillations, this would provide strong support to the hypothesis that this behavior is an emergent rather than individual based phenomenon.

The unconstrained model describes the load's motion as the result of two general forces. The first is the force that ants apply  in the short period that immediately follows their attachment to the load. These ants are well aware of their spatial position and the force they exert on the load is aimed towards the nest. Once attached to the load, informed ants lose their directional knowledge over a timescale of tens of seconds  \cite{gelblum2015ant} and become uninformed.

The forces exerted by uninformed ants contribute to the persistence of this system. These ants are unaware of the nest direction and align their pull with the current direction of the collective motion which they sense via their point of contact. This mechanical sensing is the only form of communication in this simple model. Specifically, uninformed ants switch between two possible roles: a puller ant that exerts a force, $f_0$, along its body axis, and a lifter ant that lifts the object to reduce friction.  Switching rates are:
% $\boldsymbol{\mathrm{p}}\cdot \boldsymbol{\mathrm{f}}_{\mathrm{loc}}$:
$r_{p\leftrightarrow l}\propto \exp(\pm\boldsymbol{\mathrm{p}}\cdot \boldsymbol{\mathrm{f}}_{\mathrm{loc}}/F_{\mathrm{ind}} )$, where $\boldsymbol{\mathrm{p}}$ is the orientation of the ant, $\boldsymbol{\mathrm{f}}_{\mathrm{loc}}$ the force she senses at her point of attachment, and $F_{\mathrm{ind}}$ is a parameter that is inversely proportional to the strength of the coupling between each uninformed ant and the group (in units of force. For more details see Section SI 2.2-2.3). The signs in the argument of the exponent are chosen such that ants in the leading edge of the load tend to pull, while those at the back tend to lift. \textcolor{black}{Broadly speaking, $F_{\mathrm{ind}}$ sets the scale of an ant's reaction to the force applied by all other ants, $\boldsymbol{\mathrm{f}}_{\mathrm{loc}}$. If the sensed force is greater than $F_{\mathrm{ind}}$ then the ant will tend to comply with it and pull if at the leading edge or lift otherwise. Therefore, smaller values of $F_{\mathrm{ind}}$ correspond to stronger coupling. Similarly, if $\boldsymbol{\mathrm{f}}_{\mathrm{loc}} \ll F_{\mathrm{ind}}$  the ant will switch between pulling or lifting in a manner that is independent of the forces applied by other carriers.} For strong enough coupling, $F_{ind}\lesssim F^{\mathrm{c}}_{\mathrm{ind}} = N f_0/2$, the uninformed ants spontaneously break directional symmetry such that the system becomes ordered and moves persistently at a typical velocity  \cite{gelblum2015ant}.

In the absence of obstacles the general direction of motion is towards the nest such that the forces exerted by informed ants tend to be aligned with the persistence promoting forces of the non-informed carriers.

 We next modified the model to describe cooperative transport near an obstacle. This minimal modification used Lagrangian formalism to ensure that the motion complies with the constraints imposed by the tether  (see Sections SI 2.1-2.2, and figure SI 2). Importantly, the modified model did not include any change in ant decision making. Informed ants apply forces in the direction of the nest but are completely oblivious of the constraint. Uninformed ants react to the total force on the load which, in this case, includes not only the forces exerted by all other ants but also the tension in the tether.

\begin{figure}
\includegraphics[scale=1]{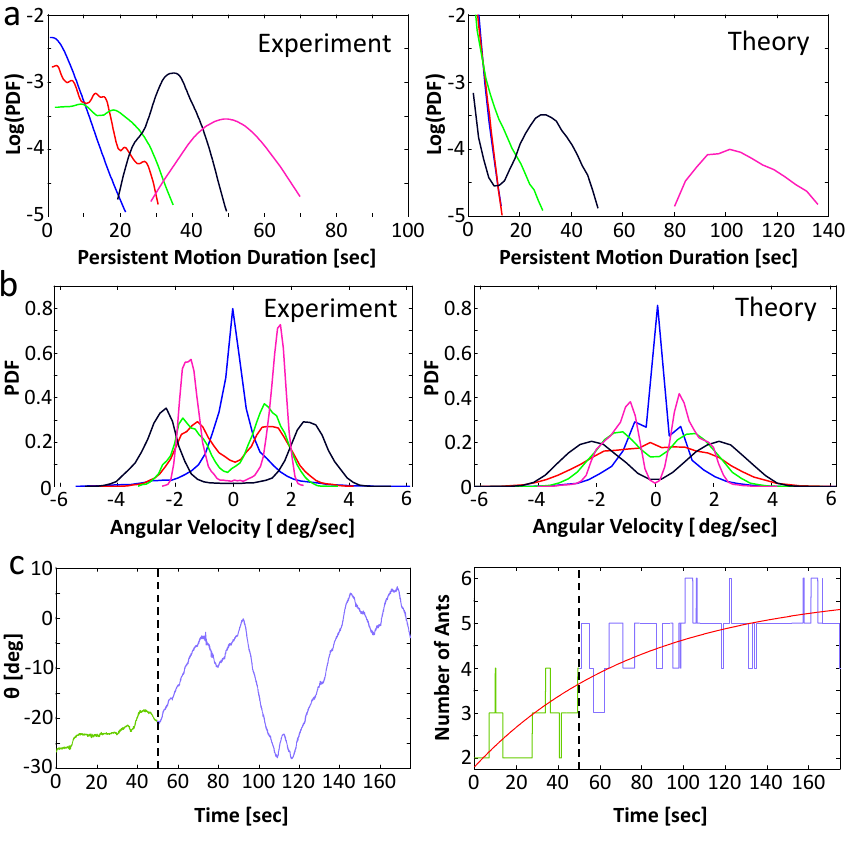}
\caption{Collective motion transitions induced by change in system size. \textbf{(a)} Distributions of persistent motion duration, defined as duration where the load moves in one direction, without stopping or turning, for different load sizes/number of ants (blue -- 0.15 cm radius 1-3 ants, red –-- 0.15 cm 3-5 ants, green --– 0.5 cm $\sim$10 ants, black –-- 1 cm $\sim$30 ants, pink –-- 4 cm $\sim$100 ants) in semilog scale, for both experiment (left) and theory (right). Note the transition from random walk distribution (characterized by exponential decay) to deterministic oscillations (characterized by a single dominant frequency). \textbf{(b)} Angular velocity distributions for different system sizes (color scheme as in (a)). Note the transition from a unimodal distribution, centered at zero, (blue) to bimodal distributions, implying persistent motion. \textcolor{black}{Data for panels (a) and (b) was collected over a total of about $5$ hours of motion.} \textbf{(c)} Left panel: An example of a transition from the random walk-like phase (green) to the persistent oscillatory phase (light purple). This transition occurs naturally as more ants join the carrying group (right panel). Red curve in right panel serves as a guide to the eye. Dashed black line depicts the estimated time point of the transition. }
\end{figure}
%fit to $a\left (1-exp(bx))  \right )+c$, inspired by the dynamics of charging a capacitor in an RC-circuit.}

 Simulating the constrained model reveals an oscillatory motion which agrees with experimental observations. We fixed the model's free parameters by fitting to several features of the system including angular velocity and amplitude distributions, power spectrum density, and phase space trajectories (see figure 2 and Section SI 2.3). \textcolor{black}{If individual ants were to directly react to the presence of a constraint (\emph{e.g.} by increasing their pulling force) then this would imply that the fitted model  parameters change in the presence of a constraint. To rule out this possibility, we conducted an experiment where the load was repeatedly released and tethered (see figure SI 3). This staggered protocol allowed us to compare measurements between periods that are proximal in time such that  all relevant conditions (\emph{e.g.} number of carriers, temperature) remain constant and the only difference is either the presence or absence of  the constraint.}
 These comparisons reveal that model parameters  are not affected by the constraint itself (see Section SI 1.2, and figure SI 3).
Finally, the fitted parameters place the system in its ordered, highly persistent, phase \cite{gelblum2015ant}.

Intuitively, the tether constraint works to decouple the forces exerted by informed ants and directed towards the nest from the persistent forces of uninformed ants which are aligned with the direction of motion which, in this case,  must be tangential. These two directionalities do not generally coincide and the inability to simultaneously satisfy the pull of informed ants with that of their uninformed counterparts leads to the observed oscillations.

We conclude that the model that was initially developed for  cooperative transport over an open area \cite{gelblum2015ant} can explain the deterministic oscillations that occur in the presence of constraints. This provides strong supporting evidence for this model and its assumptions. Importantly, it implies that the cooperative motion is not a \emph{wisdom-of-the-crowds} phenomenon \cite{galton1907vox} in which the forces of a large number of poorly informed ants average out to increase overall accuracy \cite{cronin2014ants,sasaki2013ant}. Such  averaging behavior could result in fluctuations of the load around the direction to the nest (see figure SI 4) but not in the observed, nearly-deterministic, oscillations (figures 1-2).  Rather, these results support the model's assumption that the ants divide between those that pull towards the nest (informed ants) and those that apply their forces so that they align with the current direction of motion (non-informed ants) \cite{gelblum2015ant}. Further, while the model lacks any type of mechanism by which ants can sense the obstacle, it, nevertheless, predicts the transition into an oscillatory motion near obstacles. This provides a first indication that the observed large amplitude oscillations are an emergent phenomenon that requires no behavioral change or special problem solving capabilities from individual ants.

\subsection*{Minimal group size is necessary for the emergence of oscillations}
Cooperative transport occurs for a wide range of load-sizes. Comparing the collective motion in small \textit{vs.} large systems provides a direct experimental method by which one can distinguish emergent effects from those that originate at the level of the individual. In the context of our system, this can be done by studying how oscillations relate to group size. Indeed, our model predicts that varying the size of the carrying group can lead to a finite-size order-disorder transition \cite{gelblum2015ant,amit1984field,tunstrom2013collective}).
An ordered, pendulum-like motion is expected to become evident when the system is larger than some threshold size.

We repeated the experiments with different load sizes while keeping the mass per ant roughly constant. In figure 3a, we plot the distributions of times between direction changes for several load sizes, and compare them to the corresponding simulation results. As system size increases, the distribution exhibits a clear transition in shape, shifting from an exponential decay to a distribution that is peaked at a finite time period. In accordance, angular velocity distributions change from unimodal to bimodal (figure 3b). The transition occurs, in both experiment and simulation, at a system size of around 5 ants. Indeed, at very small sizes ($1-3$ ants) we observe no oscillatory behavior whatsoever. This is direct evidence that a single ant does not engage in the ordered perpendicular motion that characterizes the group.

The disorder to order transition can be induced by varying load size but also by increasing the number of carrying ants for a fixed size (see figure SI 5a). The latter process can occur naturally with time (figure 3c):
When a tethered load is carried by a few ants it performs a random walk around the direction of the nest. This relative immobility of the team provides an opportunity for efficient recruitment of more ants. Indeed, as time passes, more and more ants join the load-carrying team (figure 3d), eventually surpassing the threshold and transitioning the group into the ordered, large amplitude phase (figure 3c). This self-organized transition can help the ants get around an obstacle, without resorting to any individual-level changes in behavioral modes.

\begin{figure}
\includegraphics[scale=1]{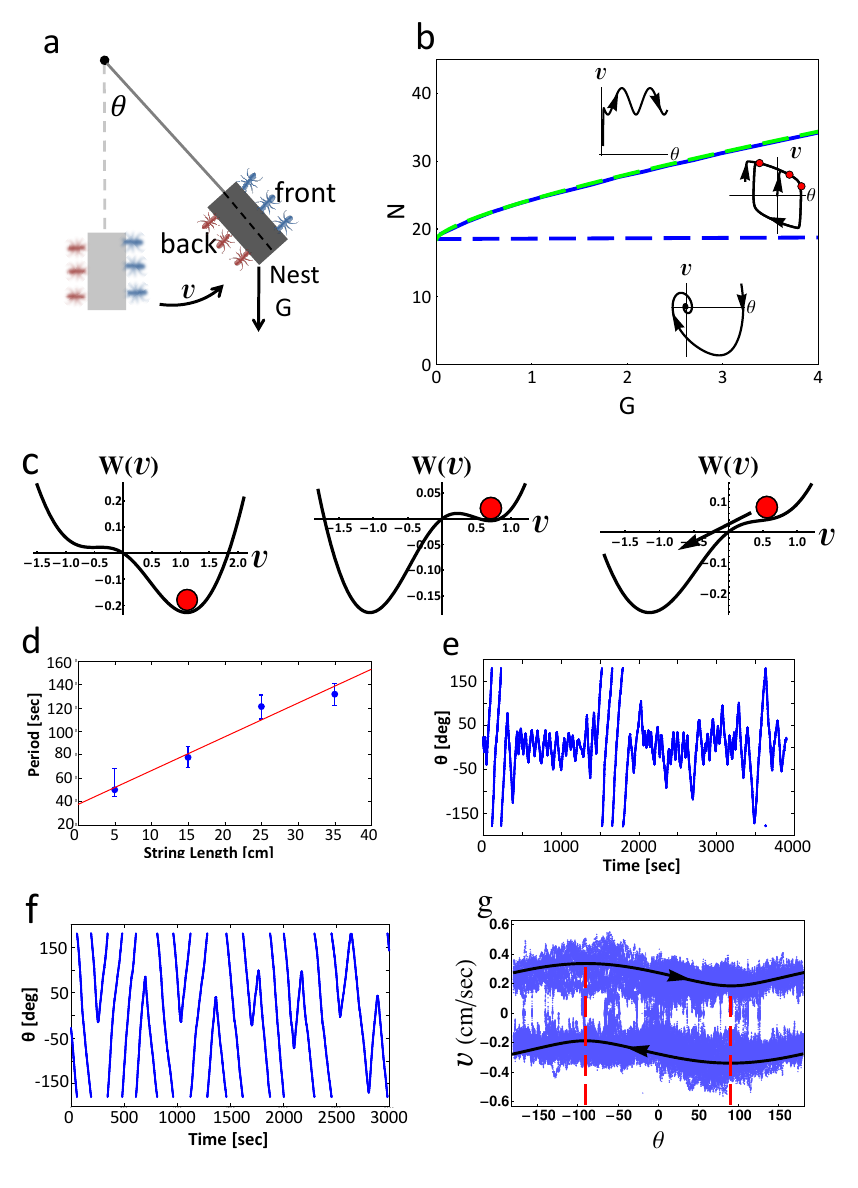}
\caption{Simplified model with three phases of motion. \textbf{(a)} Schematic depiction of the model. The object has two defined sides and can only move tangentially. {\color{black} Uninformed ants can be in either of two states: pullers, $s=+1$ (blue ants) or lifters, $s=-1$ (red ants).} Informed ants are modeled as a constant pulling force towards the nest ($G$). \textbf{(b)}  Bifurcation diagram of the simplified model, with control parameters $G$ (proportional to the number of informed ants) and $N$ (number of uninformed ants). The system exhibits three phases: stationary, oscillations, and complete rotations. Dashed blue denotes the transition between stationary and oscillatory phases, given by Eq.\ref{critical2}. The solid blue line denotes the numerical solution for the transition between the oscillatory and complete rotations phases (the dashed green line is an analytic approximation for this transition line, given by Eq.\ref{critical3}). Insets depict phase space flow diagrams for the three different phases. {\color{black}\textbf{(c)} The free energy $W(v)$ of Eq.\ref{Wv} for the red points  corresponding to $\theta = -45^o,20^o,60^o$ as marked on the limit cycle depicted in panel (b). The velocity switches its direction when the meta-stable state disappears (rightmost panel).} \textbf{(d)} Experimental oscillation period as a function of string length. Points are binned medians. Error bars are standard deviations of median distributions obtained through bootstrapping of the data. The linear trend shown (red) is reproduced by the 1D model (see SI). \textcolor{black}{Data collected over a total of about $8$ hours of motion.} \textbf{(e-f)} Examples of angular time series for 0.15 cm (e) and 1 cm (f) radius loads, held by a stiff rod. The larger load tends to perform complete rotations whereas the smaller load mainly oscillates around the direction of the nest. \textbf{(g)} Phase space of experimental complete rotations (blue points) and example trajectories from numerically solving equations 1-2 (black solid curves). Red dashed lines denote extremal angles, $\pm 90^o$ and $\theta=0$ is the nest direction. \textcolor{black}{Data represents over $3$ hours of motion.} }
\end{figure}

\subsection*{Collective modes of motion}

To study the dynamics and the bifurcation structure \cite{strogatz2014nonlinear} of the deterministic relaxation oscillations we constructed {\color{black} a simple version of our model (figure 4a, and Section SI 2.4), where the motion of the load is constrained to move on a circular circumference with radius $L$. The load geometry is taken to have only two sides: a front and a back (figure 4a). Uninformed ants are represented by $N$ spins $s=\pm 1$ ($N/2$ spins on each side), that represent the role of the ant: $s=1$ for puller and $s=-1$ for lifter. Each puller exerts a tangential force $f_0$ directed towards her side of the load. All the spins (ants) interact with all the others by sensing of the motion of the rigid load. Interactions are ferromagnetic (anti-ferromagnetic) with spins on the same (other) side, such that a puller ant in the front tends to turn other ants in the front into pullers and those in the back into lifters (figure 4a)}. Informed ants are represented by a constant force with strength $G$, which is proportional to their number and directed to the nest. The coupling between the spin dynamics and the spatial coordinate destroys detailed balance and this results in non-equilibrium  states. This stands in contrast to our description of unconstrained cooperative transport, that can be described using equilibrium Hamiltonian dynamics \cite{gelblum2015ant}. The model yields the following equations of motion for the angle and angular velocity $\left( \theta, v \right)$ (see derivation in Subsection SI 2.4.1):
\begin{eqnarray}
\frac{d\theta}{dt}&=& \frac{v}{L} \\ \nonumber
\frac{1}{k_\mathrm{c}}\frac{d v}{dt}&=& \frac{1}{\gamma k_\mathrm{c}}\frac{d f_{tot}}{dt} = \frac{1}{\gamma k_\mathrm{c}}\frac{ d\left( f_0\left( n^{front}_p - n^{back}_p\right)  - G sin\theta\right) }{dt} \label{soe2} \\
&=&-\frac{\tilde{G}}{k_\mathrm{c} L}v\cos\left(\theta \right) +f_0\tilde{N}\sinh\left( \frac{v}{\tilde{F}_{\mathrm{ind}}}\right) \\ \nonumber
&-&2\left( v +\tilde{G}\sin\left( \theta\right) \right)\cosh\left( \frac{v}{\tilde{F}_{\mathrm{ind}}}\right)
\end{eqnarray}
{\color{black} where $n^{front/back}_p$ denote the number of pullers in the front/back, $k_c$ is the rate of role changing at zero speed, $\gamma$ is the friction (mass) of the load and tilde sign denotes normalization by $\gamma$.
 The terms in Eq. \ref{soe2} are as follows: the first term signifies the (change in) force directly applied by the informed ants, this term becomes small for large tether length $L$. The second and third terms signify the changes in the persistent force that occur as uninformed ants switch between puller and lifter states.}

 %the first term is due to the change in the projection of the external force as the object is moving; the second and third terms are the result of summing all the possible events of switching between puller and lifter states at the back and at the front.}

The bifurcation diagram of the solutions to these equations in the space of numbers of uninformed and informed ants (proportional to $(N,G)$ respectively) is presented in figure 4b.
The case in which $N=0$ includes informed ants only and does not elicit any oscillations with any fluctuations around the central point diminishing for large values of $G$. This is, again, an indication that actual cooperative transport, where oscillations are more pronounced in larger systems (figure 3a), cannot arise if all carriers are (imperfectly)  informed regarding the the nest direction such that the collective direction of motion is the average of their applied forces.
The system undergoes a supercritical Hopf bifurcation at a critical number of uninformed ants $N$ (dashed horizontal blue line in figure 4b)
\begin{equation}
N_{\mathrm{c}}= \frac{F_{\mathrm{ind}}}{f_0}\left(2+\tilde{G}/k_cL\right)
\label{critical2}
\end{equation}
such that for $N>N_{\mathrm{c}}$ it exhibits a limit cycle in phase space that is similar to the experimentally observed trajectory (black lines in Figure 2b). { The parameters of the model that fit the experiments induce a separation of time-scales: during most of the oscillation the velocity changes very slowly compared to the angle, while the velocity switches sign very fast upon reaching a threshold angle. The slow evolution of $v$ during most of the cycle means that we can take $dv/dt\approx 0$ in Eq. \ref{soe2} (neglecting the small first term on the r.h.s.), and calculate the free energy landscape $W(v)$ (with external field). The stationary states are the solution of $dW(v)/dv=0$:
\begin{eqnarray} \nonumber
&W(v)&=\int\left(v-\frac{\tilde{N} f_0}{2} \tanh\left( \frac{v}{\tilde{F}_{\mathrm{ind}}}\right) +\tilde{G} \sin\left(\theta \right)\right)dv \\
&=& \hspace{-0.5cm}\frac{v^2}{2} - \frac{\tilde{N}f_0\tilde{F}_{\mathrm{ind}}}{2}\log\left(\cosh\left(\frac{v}{\tilde{F}_{\mathrm{ind}}} \right)  \right)+\tilde{G}v \sin\left(\theta \right)
\label{Wv}
\end{eqnarray}
We plot $W(v)$ for three points along the limit cycle (figure 4c) where the velocity of the system is represented by the red circle. Note that the energy landscape in this problem changes with time since it depends on the angle, which evolves due to the non-zero velocity. The leftmost panel of figure 4c depicts the part of the motion where the load moves with the external force, and therefore at the global minimum. This stable point at non-zero speed, rather than small random motion around the direction pointing to the nest ($\theta=0$), is what drives oscillations. After crossing the $\theta=0$ position, the load moves against the external force and the system is in a meta-stable state (central panel), which eventually becomes unstable and there is a fast relaxation to the global minimum as the velocity changes sign (rightmost panel). Interestingly, this direction switch does not require the attachment of a new ant to the load.

The model predicts that the oscillation period scales linearly with tether length (see Subsection SI 2.4.2) and this is indeed verified by our experimental measurements (figure 4d). The model further indicates that, in accordance with both experimental results and full model simulations (figure 3a), the period of oscillations grows with increasing system size  (see Figure SI 5b). A further prediction is that this period diverges when the system reaches an upper critical size.
Above this critical size we encounter a third dynamic regime, not observed for the string-tethered system, where the system exhibits complete rotations (over $360^o$, Figure 4b). We find an analytic approximation for the transition between oscillations and complete rotations, in terms of the critical restoring force of the informed ants (dashed green line in figure 4b, and Subsection SI 2.4.2)
\begin{eqnarray}\nonumber
G_{\mathrm{c}} &= \frac{F_{\mathrm{ind}}}{2}\sinh\left( 2 \mathrm{arccosh}\left(\sqrt{\frac{f_0 N}{2F_{\mathrm{ind}} }} \right)\right) \nonumber\\
&-{F_{\mathrm{ind}} \mathrm{arccosh}\left(\sqrt{\frac{f_0 N}{2F_{\mathrm{ind}} }}  \right)}
\label{critical3}
\end{eqnarray}
This expression is an extremely good approximation for the numerically calculated transition line (solid blue line in figure 4b). Complete rotations appear when the limit cycle is annihilated through a heteroclinic bifurcation. In the limit of large number of ants, $N>>F_{\mathrm{ind}}/f_0$ , Eq. \ref{critical3} converges to the linear relation $G_{\mathrm{c}} \approx N/2$. Thus, if the number of uninformed ants is less than twice the number of informed ants there are no complete rotations for any object size. For a full analysis of the nullclines and phase space flows refer to SI 2.4.2 and figure SI 6.

The prediction of a complete rotation phase seems to be biologically counter-intuitive, as it requires that the ants persist in motion that takes them away from the nest. To test this, we repeated our experiments, only this time attaching loads of various sizes to a light, stiff rod  rather than a string ($R=11.5$ cm). \textcolor{black}{The stiff rod simply ensures that the constraint of  constant radius  is maintained even when oscillation angles are large.} In accordance with the model's predictions, we found that the system exhibits complete rotations (see SI Movie S2), which are more robust for larger objects (figure 4e-f, figure SI 7). In both experiment and theory, the speed extrema during these cycles appear at $\pm 90^o$ which are, indeed, the locations in which the effect of the informed ants is maximized (figure 4g).

\section*{Summary}

Animal groups exhibit collective responses to external stimuli  \cite{vicsek2012collective}. In bird flocks and fish schools, such stimuli may appear in the form of pre-trained individuals \cite{couzin2011uninformed}, reactions to a predator \cite{procaccini2011propagating,ballerini2008interaction}, or interactions with a boundary \cite{czirok2000collective,buhl2006disorder,tunstrom2013collective}. In all these examples, environmental information is mediated to the group by knowledgeable individuals  \cite{vicsek2012collective, czirok2000collective, tunstrom2013collective}.
Here we have shown that when ants collectively react to a tethered load there are no such mediators. Rather, the constraint directly affects the group as a whole, transitioning it into a collective state that promotes the solution of the problem. \textcolor{black}{Such emergent responses to environmental conditions are reminiscent of those observed in the context of ant trail formation \cite{rissing1976foraging,franks1991blind} and nest construction\cite{deneubourg1995collective}.}

\textcolor{black}{We have previously modeled cooperative transport by \emph{P. longicornis} ants in an obstacle-free environment. This model assumed that the collective motion is the result of mechanical interactions between informed ants that pull the load towards the nest and uninformed ants that conform with the current direction of motion. In this work, we test this model against a different experimental scenario in which the motion is constrained. As the model was not tailored to these settings,  the current experiments provide a stringent test of its assumptions. Indeed, we found high quantitative agreement between the model's theoretical predictions and the collective motion of the ants in constrained settings.  This provides convincing evidence to our understanding of the microscopic mechanisms that drive the intriguing phenomenon of cooperative transport.}

To summarize, we have demonstrated how a simple tether transitions the frenzied motion of ants into ordered relaxation oscillations. These oscillations are a manifestation of the order-disorder transition that occurs with increasing group size. The observed deterministic motion reflects the interplay and conflict between incoming information, spatial constraints, and persistent motion. This provides a striking example of how simple physical principles can dictate the behavior of a complex biological system.

% Second, the tethered motion is shaped by a conflict between information injected by incoming ants that directs the load towards the nest and the spatial constraint (tether). These forces pull the system in opposing directions and it is this frustration that gives rise to deterministic relaxation oscillations, a collective mode never before described in the context of animal groups. It is remarkable that from the frenzied motion of the ant swarm around the food item emerge deterministic oscillations, where the animal group is enslaved to the physical rules of non-linear dynamics. Our work suggests, more generally, that combinations of information flows and constraints can give rise to a rich variety of emergent collective phases in out of equilibrium systems.
\bibliographystyle{naturemag}
\bibliography{pendulum_citations.bib}

\section*{Acknowledgments}
We thank Rinon Gal, Abhijit Ghosh, Jonathan Ron and Arik Yochelis for help and discussions and Nirit Tsori and Benjamin Sharon for technical help. O.F. is the incumbent of the Shloimo and Michla Tomarin Career Development Chair, and would like to thank ISF grant 833/15 and the  European Research Council
($ERC$) under the European Union Horizon 2020 research and innovation programme (grant agreement No 648032). N.S.G. is the incumbent of the Lee and William Abramowitz Professorial Chair of Biophysics, and would like to thank the ISF grant 580/12 for support. E.F. is the Incumbent of the Tom Beck Research Fellow Chair in Physics of Complex Systems.
This work is made possible through the historic generosity of both the Clore and the Perlman families.\newline
A. Gelblum and I. Pinkoviezky contributed equally to this work.

\end{document}